\documentclass[letterpaper, 10 pt, conference]{ieeeconf}  

\usepackage{xcolor}
\usepackage{graphicx}
\usepackage{hyperref}
\usepackage{hyperxmp}

\IEEEoverridecommandlockouts                              

\title{\LARGE \bf
Real-time Fall Prevention system for the Next-generation of Workers
}

\author{Nicholas Cartocci$^{1}$, Antonios E. Gkikakis$^{1}$, Darwin G. Caldwell$^{1}$, Jesús Ortiz$^{1}$
\thanks{$^{1}$Department of Advanced Robotics, Italian Institute of Technology (IIT), 16152 Genoa, Italy. 
{\tt\small name.surname@iit.it}}%
}

\begin{document}

\maketitle
\thispagestyle{empty}
\pagestyle{empty}

\begin{abstract}
Developing a general-purpose wearable real-time fall-detection system is still a challenging task, especially for healthy and strong subjects, such as industrial workers that work in harsh environments. In this work, we present a hybrid approach for fall detection and prevention, which uses the dynamic model of an inverted pendulum to generate simulations of falling that are then fed to a deep learning framework. The output is a signal to activate a fall mitigation mechanism when the subject is at risk of harm. The advantage of this approach is that abstracted models can be used to efficiently generate training data for thousands of different subjects with different falling initial conditions, something that is practically impossible with real experiments. This approach is suitable for a specific type of fall, where the subjects fall without changing their initial configuration significantly, and it is the first step toward a general-purpose wearable device, with the aim of reducing fall-associated injuries in industrial environments, which can improve the safety of workers.
\end{abstract}

\section{INTRODUCTION}
According to the World Health Organization (WHO), falls take the lives of 684 thousand people each year. In addition to the number of deaths, another 172 million people suffer disability each year due to falls. Over the past two decades, fall-related deaths have increased much faster than any other type of injury. This increase is due to many factors, including an aging population and urbanization patterns \cite{WHO_2021}.

Although often overlooked, the injuries caused by falls have many consequences, not only for the person who falls and those close to them but also for the healthcare system. Specifically, the total medical cost for falls of elderly people in the US was approximately \$$50$ billion. Slip-and-fall-related injuries in the workplace account for a large number of injuries in all work sectors~\cite{WHO_2021}, including office work, and are the reason for major absences from work of more than three days, especially in Small and Medium-sized Enterprises (SMEs). 

In Italy, falls from heights cause severe injuries in workers with an average duration of absence of 47 days~\cite{INAIL_2021}. The resulting loss of about 2.5 million working days in all sectors is one of the leading causes of absence from work, with obvious negative economic repercussions for the entire national production system. The compensation paid due to fall injuries amounts to more than 90 million euros (direct costs) and represents one of the first expenditure items for the Italian Workers' Compensation Authority (INAIL). Since indirect costs can be considered as a first approximation to be about three times as high as direct costs, the total costs of injuries from falls amount to more than 370 million euros per year~\cite{INAIL_2021}.  

A key aspect of developing fall prevention systems is the detection and/or prediction of the fall event. Recent developments in the field of embedded sensors and electronics allow the integration of these technologies into wearable devices. However, real-time fall detection is still challenging due to the data quality (in the presence of interference and unstructured environments), the quantity, the heterogeneity of data, and the stringent processing time requirements (the fall must be detected within a narrow time frame).   

Our work, therefore, takes the critical issues and opportunities highlighted above and involves the study and integration of Machine Learning (ML) and Deep Learning (DL) algorithms for the detection, prediction and forecasting of falls. Our approach consists of three parts: (1) detecting a fall while happening, (2) predicting when the subject might be harmed during the fall in the near future, and (3) activating the prevention mechanism to save the subject. Moreover, we introduce a new variable, in the form of a probability, which uses a-priory information (e.g., weakness, some health condition etc.) about the subject combined with biometric signals to monitor their health; for estimating the probability of an imminent fall in the future (e.g., a sudden drop of blood pressure). Most studies focus on detecting the fall after the impact occurs~\cite{Queralta_2019, Chelli_2019, Ramachandran_2020, Usmani_2021}; but in our case, we not only wish to estimate whether the subject is at risk while falling (and obviously before an impact) but also the probability of someone falling in the near future. In this paper, we present our preliminary results on detecting and predicting an occurring fall. Our approach is a hybrid approach utilizing dynamic models and deep learning algorithms for the given task. 

This research aims to find a real-time software solution that can be integrated into a plug-and-play wearable device/robot suitable for different people, regardless of age, gender, and physical structure; it should work efficiently and effectively in most environments to prevent harm to workers as a result of an impact after falling from their feet, or heights. Unlike most previous studies~\cite{Wang_2020, Saleh_2019,  Namba_2018_2,  Lord_1991}, we focus on developing a wearable for industrial workers. Particularly those who work on construction sites are subject to higher risks; they are forced into heavy and stressful repetitive operations that affect their psychological and physical health more than other categories. Moreover, most studies ignore the reaction of the subject. For example, a young and healthy subject will not fall the same way as an older person but, in many cases, may be agile enough to stop the fall by moving quickly or getting a hold of a surrounding object. This introduces false positives that cause the wrong activation of the prevention system and could dissuade the users from wearing it. So, detecting a fall is not sufficient for the purpose of our study but also activating the prevention mechanism at the appropriate time, which depends on the user's physical abilities, current mental and health state, and surroundings. 

\section{METHODOLOGY AND SENSORS}
\label{section: S3}
\begin{figure}
    \centering
    \includegraphics[width=0.9\linewidth] {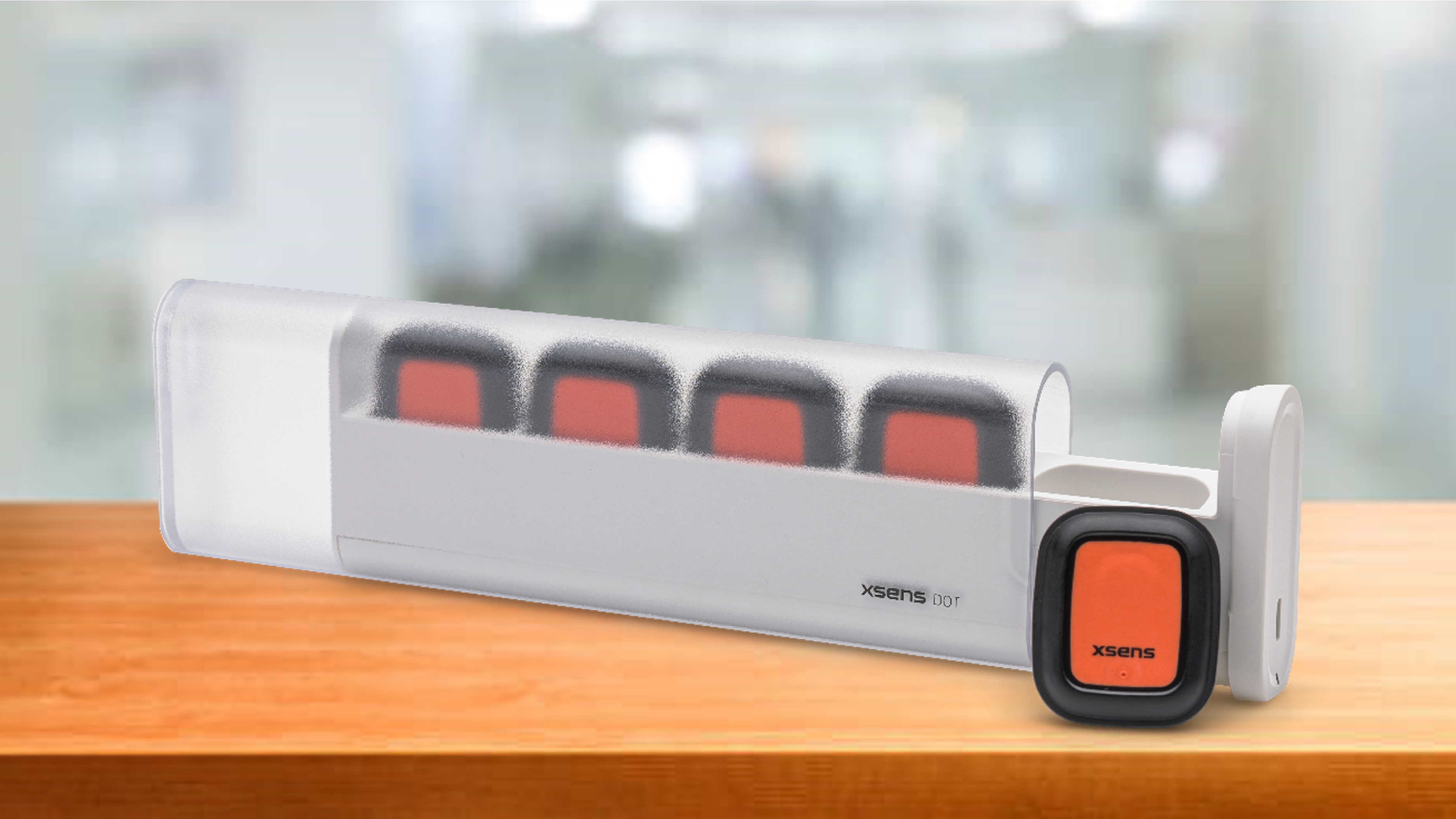}
    \caption{XSens DOT charging case and 5 IMUs.}
    \label{fig: XSens DOT}
\end{figure}
As pointed out in the previous sections, multiple types of sensors could be suitable for our objective. However, the sensors that cannot be worn due to their size, energy consumption, and computational burden are not investigated in this study since our aim is to create a wearable device for continuous and daily use. These include most vision-based sensors, which are usually placed indoors in a static and fixed manner and are also constrained by privacy issues. So, the most appropriate sensors are wearable devices that can provide inertial, physiological, and biological measurements.  

Contextually to achieve the objectives of the study, it is crucial to evaluate and compare the performance of the ML/DL techniques used with such sensors, their training, and real-time execution. In particular, DL techniques, such as Neural Networks (NNs), Convolutional Neural Networks (CNNs) and Recurrent Neural Networks (RNNs), are extensively investigated since autonomous feature extraction exploits data heterogeneity better than ML techniques~\cite{Wang_2020}.  

In the conduct of this study, a key resource is the public datasets already used for fall detection in the literature, i.e. the SisFall dataset~\cite{Sucerquia_2017}, UP-Fall~\cite{Martínez-Villaseñor_2019} and UMAFall~\cite{Casilari_2017}. With these datasets, it is possible to compare the results obtained with each sensor combination and the results obtained from state-of-the-art and proposed ML/DL techniques. However, as noted in~\cite{Wang_2020}, most public datasets utilize few sensors and contain only falls simulated by researchers. Therefore, they do not reflect all types of people and the subjects are aware that they are going to fall, affecting their behaviour. These datasets help to understand the dynamics of the fall but not to know its causes. Moreover, analyzing the datasets, we realized that the authors use sensors of limited quality in some of them. Thus, the stored data are polluted by noise and suffer from the problem of sensor drift, especially if the participants perform high-impact activities, such as running or walking fast, before simulating the fall.  

Given the complexity of the fall detection task, there is no unique and exhaustive dataset that can be used. For this reason, one of the final goals will be to build a multi-sensor dataset with as many subjects with different physical structures as possible, monitoring not only the Inertial Measurement Units (IMUs) but also the biometric sensors. 

The use of biometric signals is necessary to detect a broad spectrum of psycho-physiological effects that can be correlated to the risk of falling (e.g., heat stroke, fainting or high mental load). In addition, a pair of motion pressure insoles could be used to reconstruct the gait phases, the user’s Center of Mass (CoM), the foot Center of Pressure (CoP), and the mass distribution on each foot. In conjunction with a MoCap system, they can provide comprehensive information on the user’s balance status in static and dynamic postures.  

In the first part, we will evaluate only IMUs. For initial experiments, we plan on using the  XSens DOT~\cite{Xsens}, Fig.~\ref{fig: XSens DOT}, which is a set of 5 self-contained wireless IMUs that can be integrated into clothing. This system is flexible regarding body positioning and the number of sensors in use, so it is very handy for prototyping the final multiple IMU system.

\section{FALL DETECTION, PREDICTION AND FORECASTING}
For training ML/DL algorithms that can detect and predict falling, we need data that can capture the basic dynamics of falling and, simultaneously, they are simple enough to cover all the different variations of falling and different characteristics of humans. Although public datasets exist~\cite{Sucerquia_2017, Martínez-Villaseñor_2019, Casilari_2017}, the number of different subjects and fall scenarios is very limited, and deep learning algorithms need big data to be trained and provide generalized solutions.  
\begin{figure}
    \centering
    \includegraphics[width=0.9\linewidth] {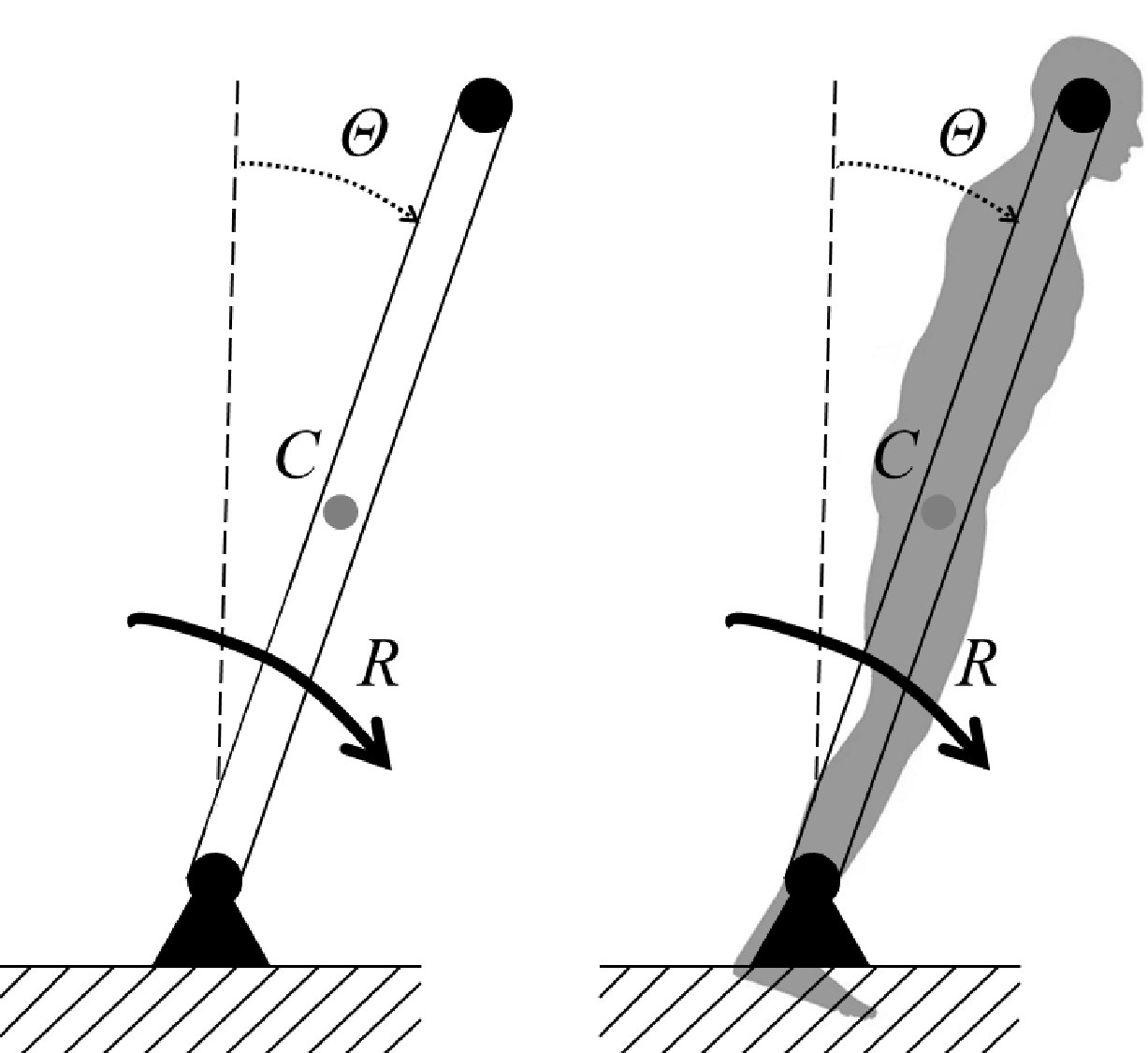}
    \caption{A human modeled as an inverted pendulum for generating simulations of falling.}
    \label{fig: inverted pendulum}
\end{figure}

\begin{figure}
    \centering
    \includegraphics[width=\linewidth] {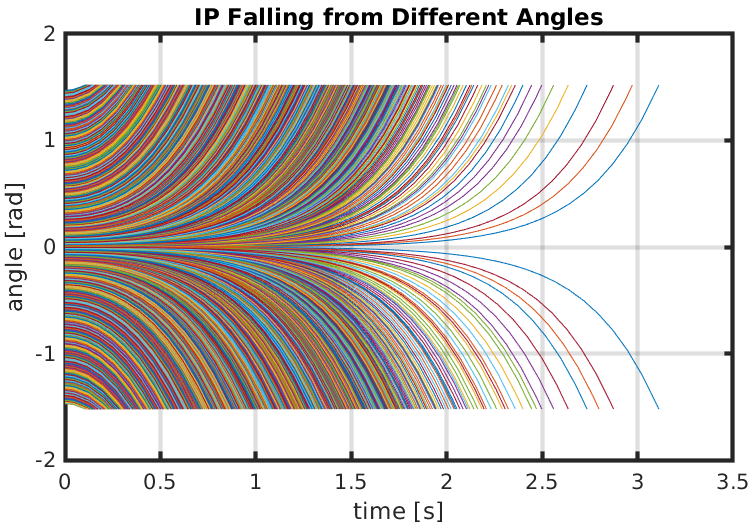}
    \caption{Example of an IP falling from different initial angles and velocities. The IP is connected to the ground via a frictionless revolute joint, and the only force acting on it is the force of gravity.}
    \label{fig: Fall Prediction}
\end{figure}
Creating a generalized model for all types of falls while taking into account all possible actions and interactions of a human is an extremely difficult task. Hence, we follow a divide-and-conquer approach, with our first sub-problem being cases where users retain their posture throughout the duration of the fall (or do not change their configuration significantly). In this type of fall, a human can be approximated as an Inverted Pendulum (IP), as presented in Figure~\ref{fig: inverted pendulum}. This type of assumption has also been used in humanoid robots for fall detection and prevention~\cite{modelabased2017}; however, in our case, we cannot obtain a detailed dynamic model of each human that uses the wearable, and for this reason, we seek a generalized real-time solution with the use of ML/DL approaches.  

This simplified model, which is computationally cheap, allows the simulation of hundreds of thousands of falls with different initial conditions and physical characteristics (height, inertia etc.) within a few hours, in contrast to computationally expensive simulations of realistic humanoid models that might require significantly more time to complete. Figure~\ref{fig: Fall Prediction} shows the evolution of the angle of an IP falling from different initial angles relative to the direction of gravity. The IP is connected to the ground via a frictionless revolute joint, and the only force acting on it is the force of gravity.

For practical purposes, one or a set of IMUs that output orientation data can be attached to a user. The number and placement of the IMUs are still subject under investigation, which will be the result of real experiments. In the dataset of~\cite{sucerquia2017sisfall}, the authors have chosen to place an IMU on the user's waist, and \cite{martinez2019up} on the left wrist, at the right pocket of the user's pants, in the middle of the waist, and under the neck. The complete framework is currently being integrated into a real-time system. 

\subsection{Fall Detection and Prediction}
\label{section: FD}
A deep Recurrent Neural Network (RNN) was chosen for fall detection, in which the input of the network is a time series composed of the relative angle of the IP and the direction of gravity, and the output is the falling probability $\mathrm{P}(\mathrm{falling})$. If $\mathrm{P}(\mathrm{falling})>0.5$, the subject is assumed to be falling. A genetic algorithm was used to identify the best network architecture by selecting the layer type between Long Short-Term Memory (LSTM), Bidirectional LSTM (BiLSTM) and Gated Recurrent Unit (GRU), and the number of hidden units in the range [30, 100] to maximize classification accuracy and recall. The optimization framework evaluated 100 different networks for 100 epochs each, and among them, a network with a hidden layer of 100 GRUs was selected. This is still incomplete work, and the next step is to measure network performance on our future multi-sensor dataset. Fall prediction will be the final part of the study since it requires the availability of experimental subjects and biometric sensors.

\subsection{Fall Forecasting}
\begin{figure}
    \centering
    \includegraphics[width=\linewidth]{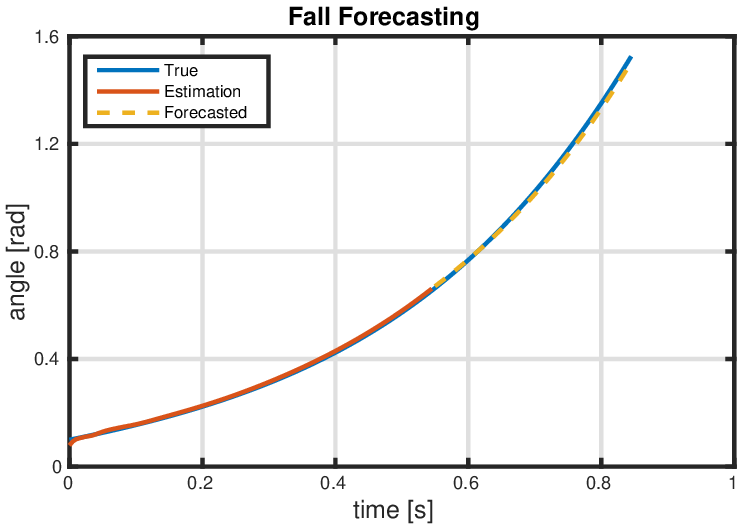}
    \caption{Example of fall forecasting of an IP falling with a non-zero initial angle and velocity.}
    \label{fig:fall_pred}
\end{figure}
For this work, we refer to fall forecasting as the forecasting of the future state of a subject that is currently falling, assuming that the subject will continue to fall. Figure~\ref{fig:fall_pred} demonstrates an example of the prediction approach in which an IP is left to fall under gravity with a small no-zero initial velocity and angle. The dashed yellow line (the prediction) almost overlaps with the future value (the blue line), and the red line is the estimation of the network with the values given up to that point. Note that this approach assumes that the user will collide with the floor at approximately $90^{\circ}$ ($1.57\ rad$). Taking into account the surroundings (e.g., a desk or a wall in the direction of the fall) would require vision-based sensors and is part of a future study.

To forecast a fall, a deep learning approach is used to extrapolate the subject's orientation into the future and identify when the subject will be at risk of harm. Similar to the approach for fall detection, a RNN was chosen for this task. The input is the same signal used for fall detection (see Section~\ref{section: FD}), but in this case, the output is a time series of the relative angle between the subject's torso and the gravity vector from an IMU (placement and number of IMUs are still under investigation). The network architecture consists of two layers of 100 GRUs each, and a simple trial-and-error approach was sufficient to generate adequate results; however, an optimization approach similar to that described in Section~\ref{section: FD} could be followed in the future, followed by our own real experiments for validating our approach.

\addtolength{\textheight}{-12cm}   





\section*{ACKNOWLEDGMENT}

This work was supported by the Italian Workers’ Compensation Authority (INAIL).



\bibliographystyle{IEEEtraN} 
\bibliography{IEEEabrv,References}

\end{document}